\documentclass[
showpacs,
preprint,
aps,
prb,
superscriptaddress,
]{revtex4-1}

\usepackage{epsfig}
\usepackage[colorlinks=false]{hyperref}
\usepackage[sort&compress]{natbib}
 \citeindextrue
 \makeindex

\newcommand{\gone}{g^{(1)}(\tau)}
\newcommand{\gtwo}{g^{(2)}(\tau)}

\newcommand{\gtwoz}{g^{(2)}(0)}

\begin{document}

\title{First- and second-order coherence properties of a quantum-dot micropillar laser}

\author{J.-S.~Tempel}
\email{jean.tempel@tu-dortmund.de} 
\author{I.\,A.~Akimov}
\author{M.~A{\ss}mann}
\affiliation{Experimentelle Physik 2, Technische Universit\"{a}t Dortmund, D-44221 Dortmund, Germany}
\author{C.~Schneider}
\author{S.~H\"ofling}
\author{C.~Kistner}
\author{S.~Reitzenstein}
\author{L.~Worschech}
\author{A.~Forchel}
\affiliation{Technische Physik, Physikalisches Institut, Wilhelm Conrad R\"{o}ntgen Research Center
for Complex Material Systems, Universit\"{a}t W\"{u}rzburg, D-97074 W\"{u}rzburg, Germany}
\author{M.~Bayer}
\affiliation{Experimentelle Physik 2, Technische Universit\"{a}t Dortmund, D-44221 Dortmund, Germany}

\date{\today}

\begin{abstract}

We present investigations on the coherence of the emission from the polarization split fundamental mode of an AlGaInAs/GaAs quantum-dot microcavity laser. Using polarization insensitive measurements of the first-order field-correlation function $g^{(1)}(\tau)$ we determine the power-dependent degree of polarization. Whereas the orthogonally-polarized components of the fundamental mode exhibit comparable strength below the lasing threshold, a degree of linear polarization of $0.99$ is observed in the coherent regime. This is also observed for increasing temperatures up to $77\,$K. Furthermore, by measuring $\gone$ for both modes separately the stronger mode is found to reveal a record coherence time of $20\,$ns. Finally, based on a theoretical model it is possible to fully characterize the cavity emission in terms of first- and second-order coherence using auxiliary data from $g^{(2)}(\tau)$ measurements performed with a recently developed streak camera technique.

\end{abstract}


										
\maketitle

\section{Introduction}

In recent years, rapid development in semiconductor technology made possible the fabrication of structured pillar microcavities, see e.g.\ \onlinecite{reitzenstein10} and references therein. Such microcavities in which localized photonic modes interact with electronic excitations, e.g.\ excitons in self-assembled quantum dots (QDs), attract a lot of attention. Due to the photonic and electronic confinement properties they are well suited candidates for studies of fundamental cavity quantum electrodynamic effects\cite{reitzenstein08,wiersig09} as well as for implementation in photonic devices.\cite{reitzenstein06,dousse09} Both, the coherence properties and the polarization stability of radiation play a decisive role with respect to laser applications and are currently under active investigation.\cite{ates07,daraei07}

In order to achieve stable lasing emission with long coherence times the use of high quality microcavities is a prerequisite. The use of self-assembled QDs as optically active material embedded between two distributed Bragg reflectors (DBRs), offering small mode volume and high quality factors as large as $Q \sim 10^{5}$ seems to be very promising.\cite{reitzenstein07} However, the nominal cylindrical symmetry of vertical-cavity surface-emitting lasers involves a degeneracy of the fundamental mode (FM) and thus potentially leads to polarization instabilities. In principle, this can be overcome by changing the pillar's cross-section to elliptical and thus lifting the degeneracy which results into two orthogonally polarized modes.\cite{gayral98,whittaker07} It is known that one of those modes can dominate the emission above the lasing threshold, revealing coherence times in the range of up to one nanosecond.\cite{ates08} In this paper we present an interferometric study of nominally cylindrical (Ga,Al)As micropillar structures. We first characterize the basic emission properties as well as the polarization behaviour. Subsequently, the micropillar's coherence properties in terms of the first- and second-order correlation function are quantified and directly measured coherence times as large as $\sim 20\,$ns are reported. Coherent lasing emission is observed over a broad range of temperatures up to $77\,$K.

\section{Experimental Section}
The studied pillar structures with diameters in the $\mu\text{m}$-range were processed from a planar microcavity designed for operation at about $900\,$nm and grown by molecular beam epitaxy.\cite{wiersig09} The DBR stacks consist of 26 upper and 33 lower pairs of alternating AlAs($74\,\text{nm}$)-GaAs($68\,\text{nm}$) layers. The central GaAs $\lambda$-cavity contains one layer of self-assembled AlGaInAs QDs with a density of about $6\cdot10^{9}\,\text{cm}^{-2}$. The sample was mounted into a He-flow cryostat with a variable temperature holder. Using a confocal optical setup equipped with a 10-times magnification microscope objective allowed for optical access to single micropillars. The pillars were excited above the GaAs bandgap and the stopband of the DBR structure by a continuous wave diode laser with an emission energy of $\hbar\omega = 1.58\,$eV. The laser beam was focused into a spot of about $10\,\mu$m in diameter. A half-wave plate in front of a Glan-Taylor prism in the detection path was used to select the desired linear polarization of the emitted signal. Our measurements were carried out in a temperature range of 10 to 77\,K.

The first-order field coherence was measured using a Michelson interferometer consisting of a 50/50 non-polarizing beam-splitter cube and one retroreflector in each of the interferometer arms. A variable path difference between the arms was realized by a mechanical translation stage providing coarse delays of up to $1.8\,\text{ns}$. The reflector in the second arm was mounted on a piezo translation stage moving on the wavelength scale. This allowed us to introduce fine delays in the range of $100\,\text{fs}$ with a resolution of about $5\,\text{as}$. The emission signal passed through the interferometer was then dispersed in a single monochromator and detected with a nitrogen-cooled charge-coupled devices camera.

\begin{figure}
	\includegraphics{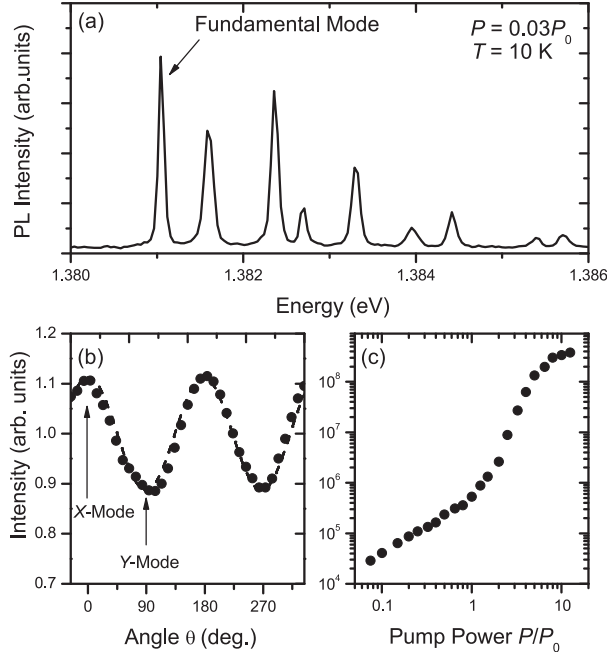}
	\caption{\label{fig:PL} (a) Photoluminescence spectrum of the $8\,\mu$m-pillar at low excitation power ($P = 0.03P_0$). The cavity's FM is indicated ($\hbar\omega=1.381\,$eV). (b) Emission intensity of the FM as a function of the polarization detection angle $\theta$. The dashed line is a sinusoidal fit. All subsequent measurements have been done at angles of $\theta=0^{\circ}$ (polarization along the $X$-axis), $90^\circ$ ($Y$-axis) and along the diagonal between both axes. (c) Input-output characteristics of the $X$-polarized component of the micropillar's FM. The smooth but distinct transition from thermal to lasing behaviour can clearly be observed.}
\end{figure}

\section{Results and Discussion}


In this paper we exemplify results obtained from the investigation of a typical single micropillar laser with a diameter of $8\,\mu$m. The spectral and polarization characteristics of the pillar's emission at low excitation power as well as an input-output curve are shown in Figure~\ref{fig:PL}. The photoluminescence spectrum taken at an excitation power of $P=0.03P_0$ (with the onset of the lasing transition region at $P_0 = 200\,\mu$W) comprises several photonic modes, which result from three dimensional optical confinement in micropillar structures.\cite{gerard96} The lowest energy peak corresponds to the resonator's FM. It features a rather narrow resolution-limited spectral linewidth of roughly $100\,\mu$eV. A plot of the integrated intensity of the fundamental cavity mode at $P=0.25P_0$ as a function of the polarization detection angle $\theta$ is shown in Fig.~\ref{fig:PL}(b) and reveals a noticeable linear polarization along $\theta \approx 0^{\circ}$. The normalized mode strength follows closely a relation $I(\theta)=1+\rho_{L}\cos(2\theta)$ with a polarization degree of $\rho_{L} = 0.12$. The characteristic direction of polarization is independent on the excitation polarization and hence defines a direction intrinsic to the given micropillar. The existence of linearly polarized emission is attributed to a deviation of the cylindrical symmetry of the micropillar's cross-section. Hereafter these main axes of the microcavity will be denoted as the $X$ ($\theta=0^{\circ}$) and $Y$ ($90^\circ$) axes. Fig.~\ref{fig:PL}(c) shows the excitation power-dependent output intensity detected in $X$ direction in a double-logarithmic plot. The typical s-shaped smooth transition from the thermal to the lasing regime can be observed, in qualitative agreement with results reported from investigations on similar micropillar structures.\cite{reitzenstein06,ates08}

\begin{figure}
\includegraphics{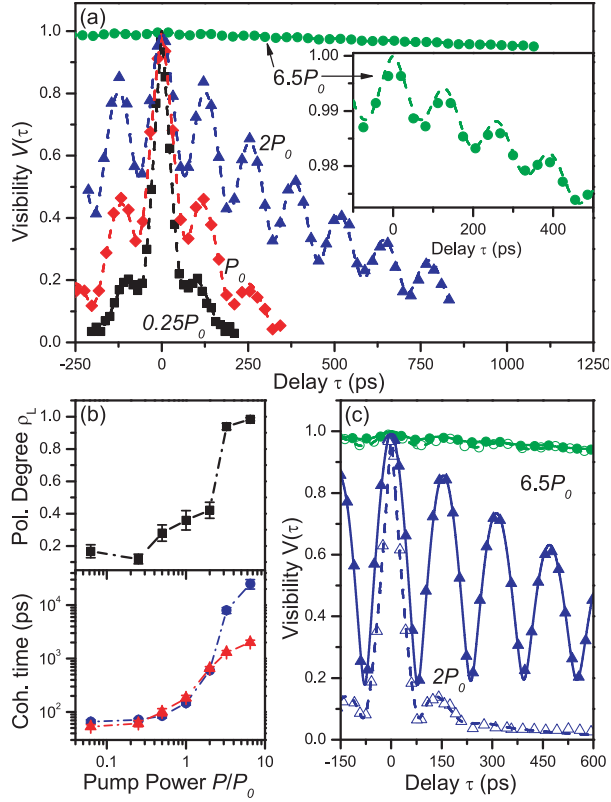}
  \caption{\label{fig:beats}
(Color online) (a) Visibility at polarization detection angle $\theta = 45^\circ$. The inset shows an enlarged visibility curve with remains of oscillatory behaviour at $P = 6.5P_0$. The curves are fits using Eq.~(\ref{eqG1osc}). (b) Pump power dependence of the polarization degree $\rho_L$ (black squares) as well as the coherence times $\tau_c^X$ (blue circles) and $\tau_c^Y$ (red triangles). (c) Visibility curves ($\theta = 45^\circ$) from another $8\,\mu$m-pillar at excitation powers of 2$P_0$ (blue triangles) and 6.5$P_0$ (green circles). Solid (open) symbols represent data taken at $10\,$K ($77\,$K).}
\end{figure}

\begin{figure}
\includegraphics{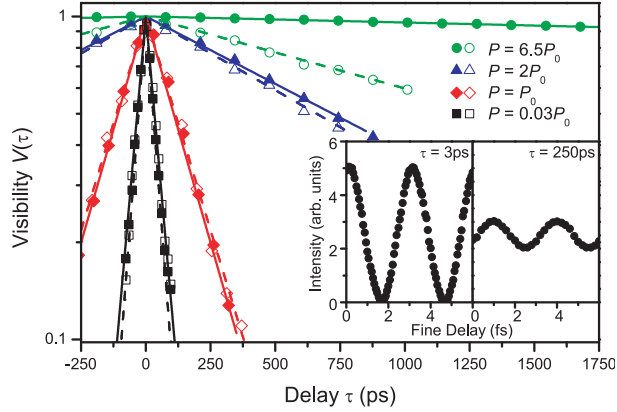}
  \caption{\label{fig:coherence}
(Color online) Visibility $V(\tau)$ of the interference fringes at different excitation powers as a function of the delay $\tau$ detected in $X$ ($\theta=0^\circ$, solid symbols) and $Y$ ($\theta=90^\circ$, open symbols) direction. The solid (dashed) lines are exponential fits to the $X$ ($Y$) data. The inset shows typical interference fringes at a delay time $\tau$ close to 0 ($V\!=0.98$) and at $250\,$ps ($V\!= 0.20$) for $P=P_0$.}
\end{figure}

Through interferometric measurement of the first-order field-correlation function $\gone$ one can determine the coherence as well as the polarization properties of the emission from the sample. The data are summarized in Figs.~\ref{fig:beats} and \ref{fig:coherence}. Examples of interference fringes observed using the Michelson interferometer at two different delays $\tau$ around 0 and 250~ps for $P=P_0$ are shown in the inset of Fig.~\ref{fig:coherence}. The signal oscillates with a period of $2\pi/\omega_0=2.995\,$fs with $\omega_0$ corresponding to the photon frequency of the cavity's FM. Using sinusoidal fits for the interference fringes at different delay times $\tau$ we evaluate the visibility $V = (I_{\text{max}}-I_{\text{min}}) / (I_{\text{max}}+I_{\text{min}})$, where $I_{\text{max}}$ and $I_{\text{min}}$ are the maximum and minimum intensity of the fringes. The dependence $V(\tau)$ is directly related to the modulus of the first-order field-correlation function of the electric field $E(t)$, $g^{(1)}(\tau) = \langle E^*(t) E(t+\tau) \rangle / \langle E^2(t) \rangle$, and, in the case of a Lorentzian power spectrum decays with a characteristic coherence time $\tau_c$.\cite{loudon73}

The relative intensity as well as the coherence times of both $X$ and $Y$ mode can directly be derived from visibility measurements at a polarization detection angle of $\theta = 45\,^{\circ}$. The resulting visibility curve is shown in Fig.~\ref{fig:beats}. In this case clear oscillations can be observed, in qualitative agreement with Ref.\ \onlinecite{ates07}. These beatings are a result of the contribution of both the $X$ and the $Y$ mode. Assuming a Lorentzian spectral shape for each of the cross-polarized modes, the visibility can be written in the form 
\begin{equation}
    V(\tau) \sim \left(\mathcal{I}^2_{X} + \mathcal{I}^2_{Y} + 2\,\mathcal{I}_{X}\,\mathcal{I}_{Y} \cos[\Delta\omega\tau]\right)^{1/2},\label{eqG1osc}
\end{equation}
where $\mathcal{I}_{i} = \alpha_i I_{i}\exp(-\vert\tau\vert/\tau_c^i)$ and $i=X,Y$ with $I_{i}$ and $\tau_c^i$ representing the intensity and the coherence time of each mode. The transmission coefficient of polarization $i$ through the optical elements is accounted for by $\alpha_i$.\footnote{The transmission ratios are $\alpha_X^{-1} = 0.51$ and $\alpha_Y^{-1} = 0.43$.} Thus, a fit to the data in Fig.~\ref{fig:beats}(a) with Eq.~(\ref{eqG1osc}) gives access to the mode splitting as well as to the coherence time and the strength of each mode.

The period of the oscillations does not depend on the pump power, but changes from one micropillar to another. In the case of the $8\,\mu$m-pillar we observe a period of roughly $120\,$ps. From this one can calculate an energy splitting between the two cross-polarized modes of $\hbar\Delta\omega = (35\pm3)\,\mu$eV, which is in agreement with high resolution measurements (not shown). The power dependence of the coherence times and the degree of polarization are shown in Fig.~\ref{fig:beats}(b). As expected, at low powers the value of $\rho_L$ evaluated from interferograms is rather small and in agreement with $\rho_L \sim 0.1$ obtained from Fig.~\ref{fig:PL}(b). In this regime both modes contribute significantly leading to polarization beats, which is undesirable for laser operation. At excitation powers above the threshold region the $X$ polarized mode dominates the emission which results in polarization stable single mode operation with $\rho_L = 0.99 \pm 0.01$.

Furthermore, while the coherence times of both modes are increasing significantly in the lasing transition region, $\tau_c^X$ increases much stronger than $\tau^Y_{c}$ and reaches a value of $(21\pm5)\,$ns at an excitation power of $6.5P_0$, as shown in Fig.~\ref{fig:beats}(b). An increase of the temperature up to $77\,$K leads only to an increase of the laser threshold value (see Fig.~\ref{fig:beats}(c)), which is related to stronger contribution of non-radiative recombination. Therefore the passage to single polarization emission cannot be attributed to tuning particular exciton transitions in resonance with the fundamental cavity mode, which is known to be very sensitive to changes of the lattice temperature. Heating effects at large excitation powers can also be excluded. These observations can be explained with the existence of an initial inequality in the mode strengths, which manifests in slightly linear polarized emission at low excitation powers. Above the laser threshold strong non-linearities lead to the predominance of this mode. The observed coherence time of $\sim 20\,$ns in the lasing regime remains almost the same at the temperature of liquid nitrogen at $P=6.5P_0$. Considering the polarization and coherence properties derived above, one can define a threshold of this QD laser where the single mode starts to dominate, namely $6.5P_0$ which is situated at the higher plateau of the $s$-shaped input-output curve.

In order to analyze the coherence properties in more detail, particular visibility measurements with the polarization detection along the main axes were performed. As mentioned above, the Lorentzian power spectrum of each mode results in a visibility curve that decays exponentially with $\tau_c$ as depicted in Fig.~\ref{fig:coherence}. Three main features can be observed here: As described above, the coherence times of both modes increase significantly, in the case of $X$ mode from $(45\pm5)\,$ps up to roughly $21\,$ns while increasing the pump power from $0.03P_0$ to $6.5P_0$. We observe no variation from exponential decay at low power, indicating the absence of inhomogeneous broadening below the laser threshold.\cite{ates08} Secondly, the above mentioned small discrepancy in the mode strengths below threshold is also reflected in the $Q$ factors. They can be derived from the lowest measured coherence times $\tau_c^X = (45\pm5)\,$ps and $\tau_c^Y = (39\pm5)\,$ps and are found to be $Q_X \sim 47\,000$ and $Q_Y \sim 40\,000$, respectively. Furthermore, the predominance of the $X$ mode at $6.5P_0$ is also greatly manifested in Fig.~\ref{fig:coherence}, as the $Y$ mode exhibits a coherence time of $(2.0\pm0.2)\,$ns which is one order of magnitude smaller than for the $X$ mode.


To complete our investigation on the coherence properties, we present a scheme to fully characterize the QD laser emission in terms of both first and second order coherence. The second-order photon correlation function $g^{(2)}(\tau)$ determines the joint probability for subsequent emission events of two photons with a relative time delay $\tau$ and thus can be referred to as the intensity autocorrelation function of the light under investigation. In the case of equal arrival time of two photons, $g^{(2)}(\tau=0)$ exhibits values of $2$ and $1$ for a system with a large number of emitters being in the thermal and in the lasing regime, respectively. However, values below 1 are classically forbidden and thus, if observed, reveal the quantum nature of a system with a small number of emitters. The latter should be the case for the high-$Q$ micropillars covered in this paper. To take the actual number $N$ of QDs contributing to the cavity emission into account, the second-order correlation function can be expressed in terms of single-emitter contributions to the emission:\cite{carmichael78}
\begin{equation}
	g^{(2)}(\tau) = \frac{g_E^{(2)}(\tau)}{N} + \left(1-\frac1{N}\right) \left[1+\chi\cdot \vert g^{(1)}(\tau)\vert^2 \right]~.\label{g2g1}
\end{equation}
Here, $g_E^{(2)}(\tau)$ is the intensity autocorrelation function of the electric field emitted by a single QD, i.e.\ as described in Ref.\ \onlinecite{michler00}. The factor $\chi$ is introduced as a fitting factor to take account for stimulated emission.

\begin{figure}
\includegraphics{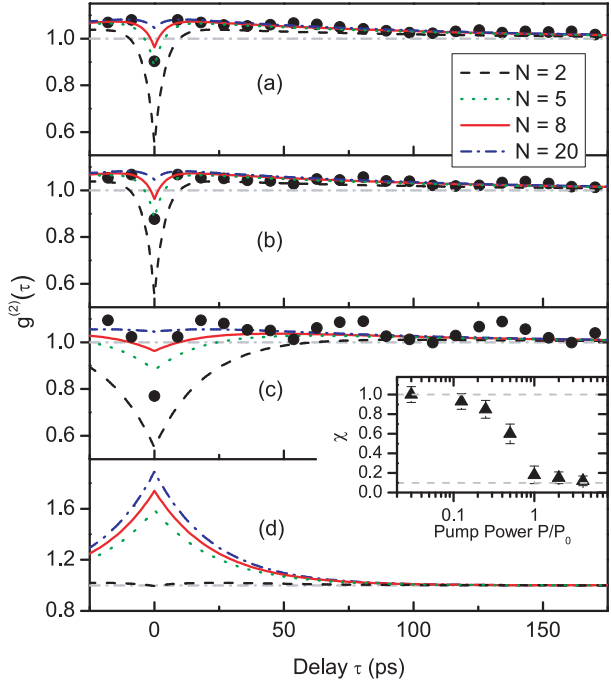}
  \caption{\label{fig:firstandsecond}
(Color online) Second-order intensity correlation function of the $8\,\mu$m pillar's FM at excitation powers of (a) $4P_0$, (b) $2P_0$, (c) $P_0$ and (d) $0.03P_0$. The figure shows both experimental results from streak camera measurements,\cite{wiersig09} as well as simulations using Eq.~(\ref{g2g1}) taking into account the coherence times derived from $\gone$ measurements shown in Fig.~\ref{fig:coherence}. At very low excitation power no measurements could be achieved due to limited sensitivity of the streak device. The inset shows the power dependence of the cavity feedback factor $\chi$.}
\end{figure}

Hence, following Eq.~(\ref{g2g1}) it is possible to compare the data obtained from $g^{(1)}(\tau)$ measurements, i.e.\ the coherence time $\tau_c$, with experimental results for $g^{(2)}(\tau)$ by choosing appropriate parameters for $N$, $\chi$ as well as for the antibunching decay rate included in $g_E^{(2)}(\tau)$. The second-order correlation function was measured using a modified streak camera equipped with an additional horizontal deflection unit. By operating in the single-photon counting mode, we are able to measure the photon statistics with a resolution down to $2\,$ps. Details of this technique were reported elsewhere.\cite{wiersig09,assmann10}

In Fig.~\ref{fig:firstandsecond}, experimental results of the second-order correlation function are shown for different excitation powers. Different $\gtwo$ curves, which have been calculated using Eq.~(\ref{g2g1}), are also shown. Thereby the coherence times derived from the $\gone$ measurements described above have been used. Also, $\chi$ has been optimized for fixed values of $N$. It should be noted that $g^{(2)}(\tau)$ was measured by pulsed laser excitation (center energy $\hbar\omega = 1.58\,$eV). Therefore, the effective coherence time of the QD-cavity system is limited by the photon lifetime which decreases from $\sim\!250\,$ps at $P_0$ to $\sim\!150\,$ps at $4P_0$.\footnote{With the photon lifetime $\tau_0$, the effective coherence time of the system is $\tau^{-1}_{c,\text{eff}} = (2\tau_0)^{-1}+\tau_c^{-1}$.} As illustrated in Fig.~\ref{fig:firstandsecond}, the calculations show a reasonable agreement with the experimental $\gtwo$ data. Furthermore, our calculations reveal the fitting factor $\chi$ to strongly depend on the excitation power. It is found to be $\sim\!1$ in the thermal regime ($0.03P_0$), in good agreement with the Siegert relation
\begin{equation}
	\gtwo = 1+|\gone|^2~.
\end{equation}
As a result of cavity feedback, $\chi$ then decreases to roughly $0.1$ in the lasing transition region at and above $P_0$, as depicted in the inset of Fig.~\ref{fig:firstandsecond}.

\begin{figure}
\includegraphics{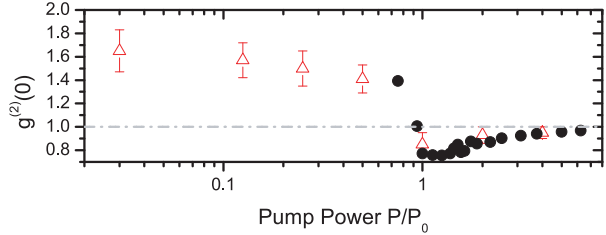}
  \caption{\label{fig:gtwozero}
(Color online) Excitation power dependence of the equal-time second-order correlation function $\gtwoz$. Data obtained from $\gtwo$ ($\gone$) measurements are shown as solid (open) symbols.}
\end{figure}

It can be seen from Fig.~\ref{fig:firstandsecond}(a)-(c) that our rather simple model fits the data best for a number of emitter of $5\leq N\leq8$, which is in agreement with the value assumed for the modeling in Ref.\ \onlinecite{wiersig09}. This highlights the fact that the high-$Q$ micropillar is indeed a few-emitter cavity system being manifested in an antibunching signal. Nevertheless, it should be mentioned that our model is only valid for cavity systems with a spontaneous emission coupling factor $\beta$ being not too high.\cite{gies07} For the investigated micropillars a value of $\beta\lesssim0.1$ is assumed. Also, it is assumed that phonon-assisted cavity feeding effects by non-resonant QDs can be neglected at this stage as they occur on a much slower time scale as compared to our experimental configuration.\cite{hohenester09}

Furthermore, as shown in Fig.~\ref{fig:firstandsecond}(d), it is possible to simulate second-order correlation values at low excitation power, as reliable $g^{(2)}(\tau)$ measurements are rather difficult in this regime due to the limited sensitivity of the streak camera in the near infrared. This way the model can also be used to derive values for equal-time correlation $g^{(2)}(0)$ as depicted in Fig.~\ref{fig:gtwozero} which underlines the good agreement of direct and indirect $\gtwo$ measurements. Thus, first-order coherence measurements can be used to extend $g^{(2)}(\tau)$ series to the previously unavailable low power domain.

\section{Conclusions}

In summary, we have investigated systematically the first and second-order coherence properties as well as the polarization behaviour of a QD micropillar laser. It has been shown that even nominally cylindrical pillars can feature polarization stable emission above the lasing threshold. As this behaviour remains unchanged at higher temperatures, it is attributed to a residual ellipticity of the pillar's cross-section. We show that measurements of the first-order correlation function can be used to extend second-order photon correlation data to low signal ranges. Additionally a record value for the coherence time of a QD laser has been found, our findings have thus various implications for laser devices operating at elevated temperatures.

\begin{acknowledgments}
The authors would like to thank M.\,M.~Glazov and D.\,R.~Yakovlev for useful discussions. This work was supported through the DFG research group ``Quantum optics in semiconductor nanostructures''.
\end{acknowledgments}

\bibliography{references}

\end{document}